\documentclass[conference]{IEEEtran}
\usepackage[T1]{fontenc}
\usepackage{bm}
\usepackage{amssymb,amsmath,amsfonts,amsthm}
\usepackage{mathrsfs}
\usepackage{cite}
\usepackage{array}
\usepackage{tabularx}
\usepackage[table]{xcolor}
\usepackage{multicol, multirow}
\usepackage{color}
\usepackage{mathtools}
\usepackage{subcaption}
\usepackage{mathtools}
\usepackage{tikz,pgf}
\usepackage{verbatim}
\usetikzlibrary{calc,positioning,mindmap,trees,decorations.pathreplacing}
\usepackage{graphicx}
\usepackage{bbm}
\usepackage[utf8]{inputenc}
\usepackage{algorithm}
\usepackage{algorithmic}

\renewcommand{\IEEEQED}{\IEEEQEDopen}

\usepackage[colorinlistoftodos,prependcaption,backgroundcolor=black!5!white,bordercolor=red]{todonotes}

\IEEEoverridecommandlockouts
\ifCLASSINFOpdf
\else
\fi
\usepackage[cmintegrals]{newtxmath}
\begin{document}
	\title{Robust and Efficient Average Consensus with  Non-Coherent Over-the-Air Aggregation}
	
	\author{\IEEEauthorblockN{ Yuhang Deng, Zheng Chen, and Erik G. Larsson} \IEEEauthorblockA{ {Department of Electrical Engineering,} {Link\"oping University}, Sweden }  \IEEEauthorblockA{E-mail: \{yuhang.deng, zheng.chen, erik.g.larsson\}@liu.se} 
	\thanks{
		This work was supported in part by Knut and Alice Wallenberg Foundation, ELLIIT, and Swedish Research Council. 
		}}
	
	\maketitle
	
	\begin{abstract}	
	Non-coherent over-the-air (OTA) computation has garnered increasing attention for its advantages in facilitating information aggregation among distributed agents in resource-constrained networks without requiring precise channel estimation. 
	A promising application scenario of this method is distributed average consensus in wireless multi-agent systems. 
	However, in such scenario, non-coherent interference from concurrent OTA transmissions can introduce bias in the consensus value. 
	To address this issue, we develop a robust distributed average consensus algorithm by formulating the consensus problem as a distributed optimization problem. 
	Using decentralized projected gradient descent (D-PGD), our proposed algorithm can achieve unbiased mean square average consensus even in the presence of non-coherent interference and noise.
	Additionally, we implement transmit power control and receive scaling mechanisms to further accelerate convergence. Simulation results demonstrate that our method can significantly enhance the convergence speed of the D-PGD algorithm for OTA average consensus without compromising accuracy.
	\end{abstract}

\section{Introduction}
\label{Introduction}

In multi-agent systems, distributed consensus algorithms allow networked agents to reach agreement on specific values of interest, which is essential for applications requiring decentralized coordination \cite{4118472ReZa}. A special case is the average consensus, where the goal is to agree on the average of agents' initial state values. The interaction protocol and network topology can greatly affect the process of reaching consensus \cite{5545370Dimakis}. The most classical interaction protocol is distributed linear iteration, where agents linearly combine their own state values with the values of their neighbors to obtain updated state values in the next iteration. The convergence conditions for achieving average consensus through distributed linear iteration are presented in \cite{200465Lin, olshevsky2009convergence}.

Earlier studies on average consensus consider either perfect communication or imperfect communication with link-level noise, packet losses or delay \cite{carli2007average, fagnani2009average}. 
In large-scale wireless multi-agent systems containing numerous agents, access control and interference management are critical for achieving efficient and accurate average consensus. Assigning orthogonal resource blocks to different nodes/links is highly inefficient for extremely large networks.
Over-the-air (OTA) computation  offers a resource-efficient solution for data aggregation \cite{10092857sahin, 10155550Chen}. Typically, the superposition property of wireless channels is considered a source of multi-user interference, which can be mitigated by orthogonal channel access or multiple antenna techniques. OTA computation leverages the signal superposition to aggregate data from distributed nodes, allowing $N$ senders to transmit simultaneously using the same frequency resources, thus increasing spectral efficiency by a factor of $N$.

Typical OTA approaches rely on coherent transmission, which requires perfect channel state information (CSI) to achieve phase alignment across different transmitters. However, in fully decentralized systems with multiple receivers, it is nearly impossible for them to achieve constructive phase-coherent combination simultaneously, even with perfect transmitter CSI. Furthermore, estimating CSI of all links in  multi-agent systems also leads to substantial signaling overhead \cite{ota-distributed-optimization}. These challenges have spurred research into non-coherent OTA aggregation \cite{michelusi2024ota,10607888Yang}.
In \cite{michelusi2024ota}, non-coherent OTA aggregation has been proposed for wireless multi-agent optimization systems employing a decentralized gradient descent (DGD) algorithm. The key idea is to use energy-based coding and decoding such that the phase information of signals is not needed for data processing.
Similarly, in \cite{10607888Yang}, a non-coherent OTA-based average consensus algorithm is studied.
In a half-duplex setting, all active receivers can receive the aggregated transmission signals from all active transmitters simultaneously. 
Following a well-designed encoding/decoding rule and consensus protocol, agents update their state values using the decoded information and the system achieves asymptotic consensus. However, the consensus value is not the exact average of the initial state values but includes a stochastic bias, which is caused by the non-coherent interference. 

In this work, we propose an average consensus algorithm based on decentralized projected gradient descent (D-PGD) with non-coherent OTA aggregation. Our proposed algorithm can reach mean square consensus on the average of the agents' initial state values without any bias. Moreover, we design a transmit power control and receive scaling method to accelerate the convergence of our algorithm. Our proposed algorithm is shown to achieve fast and robust convergence to average consensus, as validated by simulation results.

\section{Preliminaries}
\label{Preliminaries on Distributed Average Consensus}

In this section, we describe the fundamental concept of distributed average consensus, and possible mechanisms for solving the consensus problem.

\vspace{-1mm}
\subsection{Distributed Average Consensus}

\label{subsection: Average Consensus Problem}
Consider a network represented by a time-invariant undirected graph $\mathcal{G} = \left(\mathcal{N},\,\mathcal{E}\right)$, where $\mathcal{N}=\{1,\ldots, N\}$ denotes the set of agents and $\mathcal{E}\subseteq \mathcal{N}\times \mathcal{N}$ denotes the set of links. A pair of agents $i$ and $j$ is said to be connected if $(i,j)\in \mathcal{E}$.

Each agent has a real-valued state value $x_n[t]$ with $|x_n[t]| \leq  x_{\max}$, where $t \! \in \! \mathbb{N}_0$ denotes the time index. The goal of distributed consensus is for all agents to reach agreement on a certain value, $\mu$, in a distributed manner. This can be achieved by following an iteration process where agents exchange information with neighbors and update their state values in each iteration. 
Let $\mathbf{x}[t] = [x_1[t],\, x_2[t],\, \ldots,\, x_{N}[t]]^{\text{T}}$ denotes the vector of agent state values. 
The agents are considered to achieve asymptotic consensus if and only if
\begin{equation}
	\label{Eq.1}
	\lim _{t \rightarrow \infty} \mathbf{x}[t] = \mu \cdot \mathbf{1},
\vspace{-2mm}
\end{equation}
where $\mathbf{1}$ denotes the vector of all ones. A special case is that $\mu = \sum_{n\in\mathcal{N}}x_n[0] / N$, which is called the average consensus. The most commonly adopted mechanism for achieving average consensus is the distributed linear iteration written as
\begin{equation}
	\label{Eq.2}
	x_n[t+1] = \sum\limits_{m \in \mathcal{N}} w_{n,m} \cdot x_{m}[t],\ \ \forall n \in \mathcal{N},
	\vspace{-1mm}
\end{equation}
where $w_{n,m}$ denotes the weight that agent $n$ assigns to the information received from agent $m$. Due to the connectivity constraint in the network topology, we have $w_{n,m}=0$ if $n\neq m$ and $(n,m)\notin \mathcal{E}$. Let $\mathbf{W}\in \mathbb{R}^{N\times N}$ be a matrix where the $n$-th row and $m$-column is $\{w_{n,m}\}$. Then, \eqref{Eq.2} can be written in a matrix form as
\begin{equation}
	\label{Eq.3}
	\mathbf{x}[t+1]=\mathbf{W} \cdot \mathbf{x}[t].
\vspace{-1mm}
\end{equation}

The necessary and sufficient conditions for \eqref{Eq.3} to converge and reach consensus are given in \cite{200465Lin} as

(C1) $\ \mathbf{W} \cdot \mathbf{1} =\mathbf{1}$,

(C2) $\ \mathbf{1}^{\text{T}} \cdot \mathbf{W}=\mathbf{1}^{\text{T}}$,

(C3) $\ \rho\left(\mathbf{W} - \mathbf{1} \cdot \mathbf{1}^{\text{T}} / N\right)<1$,

\noindent   
where $\rho(\cdot)$ denotes the spectral radius. It is shown in \cite{200465Lin} that minimizing this spectral radius can accelerate the convergence rate of \eqref{Eq.3}.

\vspace{-1mm}
\subsection{Distributed Optimization Problem}
The average consensus problem can be equivalently reformulated as a distributed optimization problem written as
\begin{align}
	\label{Eq.4}
	\min_{|x|<x_{\max}} \ F(x) = \sum\limits_{n \in \mathcal{N}} \ F_n(x) = \sum\limits_{n \in \mathcal{N}} \ (x - x_n[0])^2 /2.
	\vspace{-1mm}
\end{align}

$F(x)$ is the global objective function, and $F_{n}(x)$ is the local objective function of $n$-th agent. $F(x)$ is a convex function since it is a linear combination of $N$ quadratic functions.  By setting $\nabla F(x) = 0$, we have the unique global optimum $x^* = \sum_{n=1}^{N}x_n[0]/N$, which is exactly equal to the average consensus value. Any distributed optimization algorithm aimed at solving \eqref{Eq.4} will eventually leads to average consensus.

We consider using the decentralized projected gradient descent (D-PGD) algorithm \cite{4749425Nedic, 5404774Nedic}. D-PGD is a simple first-order optimization method with a set constraint, relying on local gradient descent, consensus averaging, and an additional projection step. The general form of D-PGD is expressed as
\begin{equation}
	\label{Eq.5}
	\mathbf{x}[t+1]= \mathcal{P}_{\Omega} \left[\mathbf{W} \cdot \mathbf{x}[t] - \eta[t] \cdot \mathbf{d}[t]\right],
\end{equation}
where $\eta[t]$ denotes the step size and $\mathbf{d}[t]= [\nabla F_{1}(x_1[t]),$ $\nabla F_{2}(x_2[t]),\, \ldots,\, \nabla F_{N}(x_N[t])]^{\text{T}}$ is a vector containing the local gradients.  $\mathcal{P}_{\Omega}$ is the projection operator defined as
\vspace{-1mm}
\begin{equation}
	\label{Eq.projection}
	\mathcal{P}_{\Omega}[\mathbf{y}]=\arg \min _{\mathbf{x} \in \Omega}\|\mathbf{x}-\mathbf{y}\|_2,
	\vspace{-1mm}
\end{equation}
with $\Omega \! = \! [-x_{\max},\, x_{\max}]^N$ being the feasible set of $\mathbf{x}$. This projection step ensures that the updated state values are still within the feasible set. Using the objective function in \eqref{Eq.4}, we have
\vspace{-1mm}
\begin{equation}
	\label{Eq.6}
	\nabla F_{n}(x_n[t]) = x_n[t] - x_n[0], \  \forall n \in \mathcal{N}.
	\vspace{-1mm}
\end{equation}

This means that we plug in $\mathbf{d}[t] = [x_1[t] - x_1[0], x_2[t] - x_2[0], \ldots, x_N[t] - x_N[0]]^{\text{T}}$ into \eqref{Eq.5}.
Thus, we obtain the D-PGD algorithm for solving the distributed average consensus problem.

\begin{figure*}[h!]
	\centering
	\includegraphics[width=0.7\linewidth]{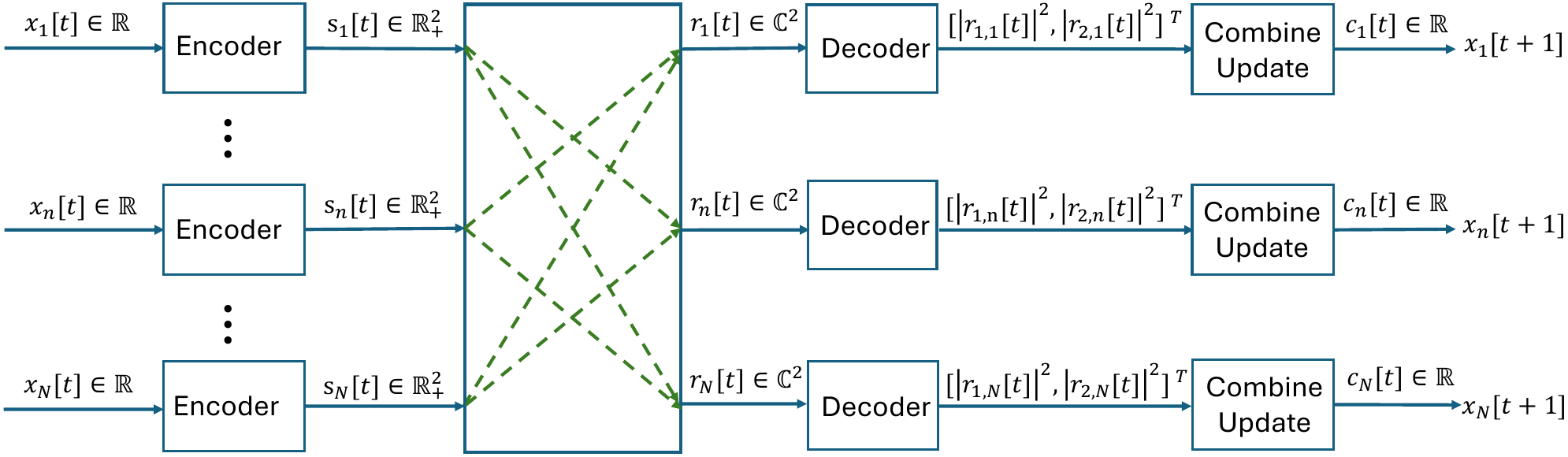}
	\caption{Computation process for distributed average consensus with non-coherent OTA aggregation over wireless fading channels.}
	\label{Fig.1:System}
	\vspace{-4mm}
\end{figure*}

\section{Non-Coherent Over-the-Air Aggregation for Distributed Consensus}
\label{Non-Coherent Over-the-Air Aggregation}

In modern engineering applications involving distributed consensus, such as swarm robotics and fleet management, the communication process is generally supported by wireless networks. Communication coordination for distributed consensus is challenging due to the ``many-to-many'' topology, where all agents must both transmit information to and receive information from their neighbors. Assigning orthogonal resources to different agents leads to inefficient utilization of communication resources in large networks. Recently, non-coherent OTA aggregation has been proposed as a resource-efficient solution. In this section, we provide a detailed description of this design and demonstrate how it can be applied to solve a distributed average consensus problem using \eqref{Eq.5}.

A wireless multi-agent system can be modeled as a complete graph with different link weights reflecting the channel quality. We assume that all agents have full-duplex capability with perfect self-interference cancellation. This means that agents can transmit and receive simultaneously without receiving any interference caused by their own transmission.\footnote{This assumption can be relaxed by assigning the nodes to two separated slots with either transmitting or receiving mode, as considered in \cite{10607888Yang}.} We denote $g_{m,n}[t]=h_{m,n}[t] \sqrt{\beta_{m,n}} \in\mathbb{C}$ as the time-varying complex channel coefficient from agent $n$ to $m$, where $\beta_{m,n}$ is the large-scale fading coefficient and $h_{m,n}[t]\sim \mathcal{CN}(0, 1)$ is the small-scale fading coefficient. We assume $h_{m,n}[t]$ is i.i.d. over time. With channel reciprocity, we have that $g_{m,n}[t]=g_{n,m}[t]$.

In Fig.~\ref{Fig.1:System}, we illustrate the entire communication and computation process of non-coherent OTA average consensus using the D-PGD algorithm in \eqref{Eq.5}. Note that the classical OTA computation typically requires instantaneous CSI to perform channel inversion such that different transmitted signals can be added with perfect phase alignment. In contrast, non-coherent OTA computation eliminates the requirement of accurate CSI; instead, the knowledge of the large-scale fading coefficients are sufficient. In the remainder of this section, we describe in detail the processing chain of our proposed design.

\vspace{-2mm}

\subsection{Precoding and signal modulation}
\label{subsection: Pre-coding and signal modulation}

The precoding block follows a modified version of the encoding design proposed in \cite{michelusi2024ota}. Let $z=[z_1,\, z_2]^{\text{T}}$ be a codebook of two codewords $z_1=r$ and $z_2=-r$, with $r \in \mathbb{R}^{+}$ and $r > x_{\max}$. Thus, any state value $x_{n}[t]$ can be represented by a convex combination of $z_1$ and $z_2$ as
\vspace{-1mm}
\begin{equation}
	\label{Eq.7}
	x_{n}[t] = p_{1,n}[t] \, z_1 + p_{2,n}[t] \, z_2, \quad \forall \, n \in \mathcal{N},\, t \in \mathbb{N}_0,
	\vspace{-1mm}
\end{equation} 
where $p_{1,n}[t],\, p_{2,n}[t] \!\in \! [0,\, 1]$, and $p_{1,n}[t]+p_{2,n}[t]=1$. Thus, $x_{n}[t]$ is encoded into a two-dimensional coefficient vector $[p_{1,n}[t] ,\, p_{2,n}[t]]^{\text{T}}$. This encoding step coverts a real-valued (possible negative) scalar into a positive vector, which is essential for the next operation. We define $\bm{\alpha} = [\alpha_{1},\, \alpha_{2},\, \ldots,\, \alpha_{N}]^{\text{T}}$ as a transmit power control vector, where $\alpha_{n} > 0$ is the power control factor of agent $n$. The transmitted symbol generated by agent $n$ in the $t$-th iteration is a two-dimensional vector:
\vspace{-1mm}
\begin{equation}
	\label{Eq.8}
s_n[t]=[s_{1,n}[t],\, s_{2,n}[t]]^{\text{T}} = [\sqrt{\alpha_{n} p_{1,n}[t]},\sqrt{\alpha_{n} p_{2,n}[t]}]^{\text{T}}.
\vspace{-1mm}
\end{equation} 

Note that the power control step is not included in the original design proposed in \cite{michelusi2024ota}, which is one of the novel aspects of this work.

\vspace{-2mm}
\subsection{Signal superposition in wireless channels} 
While all agents transmit their precoded symbols to their neighbors, they also receive the aggregated signals from their neighbors simultaneously. With full duplex capability and perfect self-interference cancellation,  the received superimposed signal at any agent $n$ in iteration $t$ is a complex-valued two-dimensional vector $r_n[t]=[r_{1,n}[t],\, r_{2,n}[t]]^{\text{T}}$, where 
\vspace{-1mm}
\begin{equation}
\label{Eq.9}	
r_{i,n}[t] = \sum_{m=1, m\neq n}^{N} g_{n,m}[t] s_{i,m}[t] + n[t], \quad  \forall i \in \{ 1,\, 2\}.
\vspace{-1mm}
\end{equation}
Here, $n[t]$  denotes the complex-valued additive white Gaussian noise (AWGN) following $n[t]\sim \mathcal{CN}(0, \sigma^2)$. 

\vspace{-2mm}
\subsection{Energy-based decoding}
With the received vector symbol $r_n[t]$, agent $n$ performs energy-based decoding by calculating the squared magnitude of each signal element in $r_n[t]$.
For each $i \in \{1,\, 2\}$, we obtain

\begin{small}
\begin{align}
	\label{Eq.10}	
	& |r_{i,n}[t]|^{2} \nonumber\\
	& =\sum_{m=1 , m\neq n}^{N}\beta_{n,m} \cdot |h_{n,m}[t]|^{2} \cdot \alpha_m \cdot p_{i,m}[t] + N_i[t] \\ 
	& + \!\! \sum\limits_{\substack{m=1 \\ m\neq n}}^{N}  \sum\limits_{\substack{k=1 \\ k\neq m,n}}^{N} \!\!\! \text{Re} \! \left(h_{n,m}^{}[t] h_{n,k}^{*}[t]\right) \sqrt{ \beta_{n,m} \alpha_m p_{i,m}[t] \cdot \beta_{n,k} \alpha_k p_{i,k}[t]}, \nonumber
\end{align}
\end{small}

\noindent
where  $N_i[t] \!=\! |n[t]|^{2} \!+ \! 2 \text{Re}\left(\sum_{m=1, m\neq n}^{N} g_{n,m}[t] \alpha_m p_{i,m}[t] n^{*}[t]\right)$. We view $N_i[t]$ as the noisy component in the decoded signal caused by the channel noise. It is easy to see that $\mathbb{E}[N_i[t]] = \sigma^2$, thanks to the independence between $g_{n,m}[t]$ and $n[t]$.

\subsection{Combining and updating}
In the final step, each agent combines the squared magnitude of the received signal and its own state value using the following combining rule:
\begin{equation}
	\label{Eq.11}
		c_n[t] = \sum\nolimits_{i=1}^{2} (	|r_{i,n}[t]|^2 - \sigma^2 ) \left(z_i-x_n[t]\right).
		\vspace{-1mm}
\end{equation}
After some intermediate steps, we obtain
\vspace{-1mm}
	\begin{align}
	\label{Eq.12}
		c_n[t] 
		=& \sum\limits_{\substack{m=1 \\ m\neq n}}^{N} |h_{n,m}[t]|^{2} \beta_{n,m} \alpha_m  \left( p_{1,m}[t]z_1 + p_{2,m}[t]z_2 - x_{n}[t] \right)  \nonumber \\
		+& \!\! \sum\limits_{\substack{m=1 \\ m\neq n}}^{N} \sum\limits_{\substack{k=1 \\ k\neq m,n}}^{N} \!\!\! \text{Re}\left(h_{n,m}^{}[t] h_{n,k}^{*}[t]\right) \sqrt{\beta_{n,m}\beta_{n,k}\alpha_m \alpha_k} \, \varepsilon_{m,k}[t], \nonumber\\
		+& \sum\nolimits_{i=1}^{2}(N_i[t] -\sigma^2) \left(z_i - x_n[t]\right),
		\vspace{-2mm}
	\end{align}
\noindent
where $\varepsilon_{m,k}[t] = \sum_{i=1}^{2}\sqrt{p_{i,m}[t] p_{i,k}[t]}\left(z_i - x_{n}[t]\right)$ is the error term caused by non-coherent interference. As shown in \eqref{Eq.12}, the combining symbol $c_n[t]$ consists of three different terms: the desired signal, the non-coherent interference from other nodes, and the aggregated noise. 

Due to all sources of randomness in the iterative computation procedure, $\{c_n[t]\}_{t=0,1,2,\ldots}$ is a time-dependent stochastic process. Let $\mathcal{F}_t$ denote the $\sigma$-algebra generated by the signals until iteration index $t-1$. In the next step, we will characterize the conditional expectation of $c_n[t]$ given $\mathcal{F}_t$, where the expectation is taken over the randomness in the $t$-th iteration. 

\begin{itemize}
	\item Let $c_{1,n}[t]$ be the first term of $c_n[t]$. According to the precoding rule in Section \ref{subsection: Pre-coding and signal modulation}, we have $p_{1,m}[t]z_1 + p_{2,m}[t]z_2 = x_m[t]$. Thus,
	\vspace{-1mm}
	\begin{equation}
		\label{Eq.13}	
		c_{1,n}[t] =\!\!\!\! \sum_{m=1 , m\neq n}^{N} \!\!\!\!|h_{n,m}[t]|^{2} \beta_{n,m} \alpha_m  \left( x_m[t] - x_{n}[t] \right),
		\vspace{-1mm}
	\end{equation}
	which represents a weighted sum of the differences between the state value of agent $n$ and those of its neighbors.
	
	\item Let $c_{2,n}[t]$ be the second term of $c_{n}[t]$. Given that $h_{n,m}[t]$ is zero-mean and i.i.d. for all $n,m \in \mathcal{N}$ and $n\neq m$, we have $\mathbb{E}[c_{2,n}[t]\vert \mathcal{F}_t] = 0$. 
	Note that $\varepsilon_{m,k}[t]$ diminishes as the state values $x_m[t]$, $x_k[t]$, and $x_n[t]$ become closer to each other, implying that $c_{2,n}[t]$ will vanish when the system approaches consensus. 
	
	\item Let $c_{3,n}[t]$ be the third term (aggregated noise) of $c_{n}[t]$.
	Since $\mathbb{E}[N_i[t]] = \sigma^2$, we have $\mathbb{E}[c_{3,n}[t] \vert \mathcal{F}_t] = 0$. 
	
\end{itemize}

Finally, we have
\vspace{-2mm}
\begin{equation}
	\label{Eq.14}	
	\mathbb{E}[c_n[t] \vert \mathcal{F}_t]=\sum\limits_{m=1, m\neq n}^{N}\beta_{m,n} \cdot \alpha_{m} \left(x_m[t]-x_n[t]\right).  
		\vspace{-1mm} 
\end{equation}

Apparently, $c_n[t]$ for all $n \in \mathcal{N}$ is an unbiased estimator of the weighted sum of the pairwise differences between agents' state values. We define $\bm{\gamma} = [\gamma_{1},\, \gamma_{2},\, \ldots,\, \gamma_{N}]^{\text{T}}$ as a receive scaling vector where $\gamma_{n} > 0,\, \forall n \in \mathcal {N}$. The state value $x_n[t]$ is updated by the following iteration rule:
\vspace{-1mm}
\begin{equation}
	\label{Eq.15}	
	x_{n}[t+1]=\mathcal{P}_{\Omega_n}\left[x_n[t]+ \zeta[t] \gamma_n c_n[t] - \eta[t] d_n[t]\right], \ \forall n \in \mathcal{N},
\end{equation}

\noindent
where $\Omega_n = [-x_{\max},\, x_{\max}]$, $d_n[t] = x_n[t]- x_n[0]$, $\zeta[t]$ is the step size of updating $x_n[t]$ with received neighbors' contribution $c_n[t]$, and $\eta[t]$ is the step size of gradient descent. We can re-write \eqref{Eq.15} in a matrix form as
\begin{equation}
	\label{Eq.16}
	\mathbf{x}[t+1]= \mathcal{P}_{\Omega}[\mathbf{W}[t] \cdot \mathbf{x}[t] - \eta[t] \cdot \mathbf{d}[t]],
\end{equation}
\noindent
where $\mathbf{W}[t] \in \mathbb{R}^{N\times N} $ is a time-varying matrix with the $(n,m)$-th element given as
	\begin{align}
		\label{Eq.17}
		& \mathbf{w}_{n, m}[t] \\
		& = \begin{cases}
			\zeta[t] \gamma_n |g_{n,m}[t]|^2 \alpha_m, &n \neq m.  \\
			1 - \zeta[t] \gamma_n \left(\sum\limits_{\substack{k=1 \\ k\neq n}}^{N}  |g_{n,k}[t]|^2 \alpha_k + \frac{c_{2,n}[t]}{x_{n}[t]} + \frac{c_{3,n}[t]}{x_{n}[t]}\right), & \text{else}.
			\end{cases} \nonumber
	\end{align}

Observing that \eqref{Eq.16} aligns with the D-PGD algorithm depicted in \eqref{Eq.5}, we name \eqref{Eq.16} as the D-PGD-based average consensus algorithm.

Note that \cite{10607888Yang} uses the primal average consensus protocol, where $\mathbf{x}[t]$ is updated according to the following rule:
\begin{equation}
	\label{Eq.18}
	\mathbf{x}[t+1]=\mathbf{W}[t] \cdot \mathbf{x}[t].
\end{equation}

Compared with the average consensus algorithm \eqref{Eq.18} proposed in \cite{10607888Yang}, our design can achieve robust consensus thanks to the additional gradient vector $ \mathbf{d}[t]$ working as a regularization term in \eqref{Eq.16}. In contrast, \eqref{Eq.18} is affected by interference and noise, preventing it from accurately converging to $x^*$. Specifically, due to the presence of $c_{2,n}[t]$ and $c_{3,n}[t]$, the mixing matrix $\mathbf{W}[t]$ is not column stochastic, resulting that $\mathbf{1}^{\text{T}} \mathbf{x}[t+1] = \mathbf{1}^{\text{T}} \mathbf{W}[t] \mathbf{x}[t] \neq \mathbf{1}^{\text{T}} \mathbf{x}[t]$. Thus, the network cannot preserve $x^* = \mathbf{1}^{\text{T}} \mathbf{x}[0] $ with the iterative update process of \eqref{Eq.18}.

\subsection{Discussions on convergence to average consensus}

Due to the persistent effect of non-coherent interference and noise, the iterate $x_n[t]$ cannot converge exactly to $x^*$. Therefore, we consider a relaxed form of average consensus called \textbf{mean-square average consensus} (MSAC), where $x_n[t]$ converges to an unbiased estimate of $x^*$:
\vspace{-1mm}
\begin{equation}
	\label{Eq.19}
	\lim _{t \rightarrow \infty} \mathbb{E}\left[\left(x_n[t]-x^*\right)^2\right]=0, \quad \forall n \in \mathcal{N}.
	\vspace{-1mm}
\end{equation}

When $\zeta[t]$ and $\eta[t]$ are decreasing functions that satisfy $\lim _{t \rightarrow \infty} \zeta[t]=0$ and $\lim _{t \rightarrow \infty} \eta[t]=0$, the D-PGD-based average consensus algorithm \eqref{Eq.16} converges to MASC. This conclusion is a corollary of \cite[Eq.37]{michelusi2024ota}, which shows that the expected square of the Euclidean distance between $x_n[t]$ and $x^*$ converge to $0$ as $t \rightarrow \infty$. Due to the space limit, a rigorous proof will be provided in an extended version of this paper.

\section{Power Control and Signal Scaling (PCSS)}
\label{Power Control and Signal Scaling (PCSS)}

In this paper, we are interested in the convergence rate of \eqref{Eq.16} and \eqref{Eq.18} for reaching consensus. According to \cite{200465Lin}, the spectral radius $\rho\left(\mathbf{W} - \mathbf{1}\mathbf{1}^{\text{T}} / N\right)$ determines the convergence rate of \eqref{Eq.3} for time-invariant mixing matrix $\mathbf{W}$. However, the mixing matrix $\mathbf{W}[t]$ in \eqref{Eq.16} and \eqref{Eq.18} are time-varying. The expectation of the mixing matrix $\overline{\mathbf{W}} = \mathbb{E}[\mathbf{W}[t] \vert \mathcal{F}_t] \big\vert_{\zeta[t] = 1}$ for all $t \in \mathbb{N}_0$ can be obtained as
\begin{align}
	\label{Eq.21}
	\big[\overline{\mathbf{W}}\big]_{n,m}= \begin{cases}
		\gamma_{n} \beta_{n,m} \alpha_{m}, & \text{if } n \neq m.  \vspace{1mm}\\
		1 - \gamma_{n} \sum_{k=1, k\neq n}^{N} \beta_{n,k} \alpha_{k}, & \text{else}.
	\end{cases}
	\vspace{-3mm} 
\end{align}

Here, we choose the step size $\zeta[t] = 1$ for faster mixing. Note that $\overline{\mathbf{W}}$ is time-invariant and satisfies the condition (C1). Additionally, $\rho(\overline{\mathbf{W}} - \mathbf{1} \mathbf{1}^{\text{T}}/N) \! \leq \! \|\overline{\mathbf{W}} - \mathbf{1} \mathbf{1}^{\text{T}}/N\|_2$, where $\|\cdot\|_2$ is the spectral norm. Therefore, to accelerate the average convergence rate, we aim to minimize the spectral norm $\| \overline{\mathbf{W}} - \mathbf{1} \mathbf{1}^{\text{T}}/N \|_2$ by designing the transmit power control vector  $\bm{\alpha}$ and receive scaling vector $\bm{\gamma}$ jointly. The optimization problem is defined as follows:
\begin{align}
	\label{Eq.22}	
	\min_{\bm{\alpha},\, \bm{\gamma}} \quad & \left\| \ \overline{\mathbf{W}} - \mathbf{1} \mathbf{1}^{\text{T}} /N \ \right\|_2, \\
	\text { s.t. } 			\quad & \alpha_n > 0,\ \alpha_n \leq \alpha_{\max}, \ \forall n\in \mathcal{N};  \nonumber \\
	& \gamma_n > 0,\ \gamma_n \leq \gamma_{\max}, \ \forall n\in \mathcal{N};  \nonumber \\
	& \gamma_{n} \sum\nolimits_{k=1, k\neq n}^{N} \beta_{n,k} \alpha_{k} = \alpha_{n} \sum\nolimits_{j=1, j\neq n}^{N} \beta_{j,n} \gamma_{j},\ \forall n\in \mathcal{N}.  \nonumber
\end{align}
where $\alpha_{\max}$ is the maximum power control coefficient, and $\gamma_{\max}$ is the maximum scaling coefficient. The additional constraint $\gamma_{n} \sum\nolimits_{k=1, k\neq n}^{N} \beta_{n,k} \alpha_{k} = \alpha_{n} \sum\nolimits_{j=1, j\neq n}^{N} \beta_{j,n} \gamma_{j}$ is imposed to ensure that $\overline{\mathbf{W}}$ is column stochastic, hence satisfying the condition (C2). 
Note that for any real-valued matrix $\mathbf{S}$, we have $\|\mathbf{S}\|_2 \leq s$ if and only if $\mathbf{S}^{\text{T}} \mathbf{S}  \preceq s^2 I$ and $s \geq 0$. Thus, we can utilize the Schur complement and rewrite the problem \eqref{Eq.22} into an equivalent semidefinite programming (SDP) problem as follows \cite{boyd2004convex}:

\begin{align}
	\label{Eq.23}	
	\min_{\bm{\alpha},\, \bm{\gamma}} \quad & \quad\quad s, \\
	\text { s.t. } 			\quad & \begin{bmatrix} s \cdot I & \overline{\mathbf{W}} - \mathbf{1} \mathbf{1}^{\text{T}}/{N} \\ \overline{\mathbf{W}}^{\text{T}} - \mathbf{1} \mathbf{1}^{\text{T}}/{N} & s \cdot I \end{bmatrix}   \succeq 0;  \nonumber \\
	 							  & \alpha_n > 0,\ \alpha_n \leq \alpha_{\max}, \ \forall n\in \mathcal{N};  \nonumber \\
								  & \gamma_n > 0,\ \gamma_n \leq \gamma_{\max}, \ \forall n\in \mathcal{N};  \nonumber \\
								  & \gamma_{n} \sum\nolimits_{k=1, k\neq n}^{N} \beta_{n,k} \alpha_{k} = \alpha_{n} \sum\nolimits_{j=1, j\neq n}^{N} \beta_{j,n} \gamma_{j},\ \forall n\in \mathcal{N}.  \nonumber
\end{align}

The SDP formulation in \eqref{Eq.23} is challenging to solve directly since the product of $\bm{\alpha}$ and $\bm{\gamma}$ is involved in the constraints. Therefore, we apply the alternating minimization (AM) algorithm \cite{ortega2000iterative} to solve it. Specifically, the AM algorithm follows an iterative procedure that alternates between optimizing one variable while keeping the other fixed, until a certain stopping criterion is reached. Detailed descriptions of the AM algorithm are provided in Algorithm \ref{Alternating Minimization}.

\vspace{-2mm}
\begin{algorithm}[h!]
	\caption{Alternating Minimization Algorithm}
	\small
	\begin{algorithmic}[1]
		\label{Alternating Minimization}
		\STATE \textbf{Initialize:} $\epsilon$, $N$, $\overline{\mathbf{W}}$, $\{ \beta_{m,n} \}_{m,n \in \mathcal{N}, \, m\neq n}$, $\bm{\alpha}[0] = [\alpha_1[0], \alpha_{2}[0],$ $ \ldots, \alpha_n[0]]^{\text{T}}$, $\bm{\gamma}[0] = [\gamma_1[0], \gamma_{2}[0], \ldots,\gamma_n[0]]^{\text{T}}$, and $k = 0$.
		\STATE $d_{\alpha} = \epsilon +1 $ and $d_{\gamma} = \epsilon +1 $.
		\WHILE {$d_{\alpha} \geq \epsilon $ or $d_{\gamma} \geq \epsilon $}
		\STATE $k=k+1$.
		\STATE Solve \eqref{Eq.23} and obtain $\bm{\alpha}[k]$ using $\bm{\gamma} = \bm{\gamma}[k-1]$.
		\STATE Solve \eqref{Eq.23} and obtain $\bm{\gamma}[k]$ using $\bm{\alpha} = \bm{\alpha}[k]$.
		\STATE $d_{\alpha} = \|\bm{\alpha}[k] - \bm{\alpha}[k-1]\|_{\infty} $ and $d_{\gamma} = \|\bm{\gamma}[k] - \bm{\gamma}[k-1]\|_{\infty} $.
		\ENDWHILE
		\STATE \textbf{Return}: $\bm{\alpha}[k]$ and $\bm{\gamma}[k]$.
	\end{algorithmic}
\end{algorithm}

Note that in \textit{step} $5$ (or $6$) of the Algorithm \ref{Alternating Minimization}, we fix $\bm{\gamma}$ (or $\bm{\alpha}$) such that the problem \eqref{Eq.23} becomes a convex minimization problem, for which the global and unique optimal solution $\bm{\alpha}$ (or $\bm{\gamma}$) can be numerically obtained in polynomial time.

\section{Simulation Results}
\label{Simulation Results}

We consider a rectangular-shaped network area of $300 \times 300\,m^2$, with $N=9$ points uniformly and randomly located inside, while ensuring a minimum distance of $d_{1}=20\,m$ between any pair of points. We assume an urban environment and generate the large-scale fading coefficient $\beta_{m, n},\, \forall m, n \!\in \!\mathcal{N}, m \!\neq\! n$ using the log-distance path loss model from \cite{goldsmith2005wireless}: 
\begin{align}
	\label{Eq.24}	
	\beta_{m, n}(\mathrm{dB}) = K(\mathrm{dB})-10 \delta \log _{10}\left[d_{m,n}/d_0\right] + \Psi(\mathrm{dB}),
	\vspace{-5mm}
\end{align}
where $K = 3 \mathrm{~dB}$ is the antenna gain parameter; $d_{m,n}$ is the distance between the agent $m$ and $n$; $d_0=10 \,m$ is the reference distance for the antenna far-field region; $\delta = 4$ is the path loss exponent; $\Psi \sim \mathcal{N}(0, \sigma_{\psi}^2)$ is a random variable with $\sigma_{\psi} = 7 \mathrm{~dB}$ representing the impact of shadowing. 

We randomly generate the initial state values $x_{n}[0]$ within $\left[-250, \, 250\right]$ for all $n \in \mathcal{N}$ agents following a uniform distribution, and implement our proposed D-PGD-based average consensus algorithm with power control and signal scaling (D-PGD-AC-PCSS), with $\alpha_{\max} = \gamma_{\max} = 5$. For performance comparison, we implement three reference schemes: $1)$ primal average consensus (AC) defined by \eqref{Eq.18}, $2)$ average consensus with power control and signal scaling (AC-PCSS), $3)$ D-PGD-based average consensus (D-PGD-AC) defined by \eqref{Eq.16}. 

Note that, for both AC and D-PGD-AC, we choose the same power control coefficient $\alpha_n = \sum_{n \in \mathcal{N}}\alpha_n^* / N$ for all $n \in \mathcal{N}$, where $\alpha_n^*$ is the optimized power control coefficient when using our PCSS design. This guarantees that all schemes (with or without power control) have the same average power consumption. In addition, $\bm{\gamma}$ is chosen to satisfy that the diagonal elements of $\overline{\mathbf{W}}$ are non-negative, which follows the design proposed in \cite{michelusi2024ota}.

Both D-PGD-AC and D-PGD-AC-PCSS use the same step-size functions $\eta[t]= t^{-1} $ and $\zeta[t]= t^{-1/10}$, hence fulfilling the requirement of decreasing and converging to $0$. In each iteration, we randomly generate the small-scale fading coefficients $h_{m, n}[t], \, \forall m, n \! \in\! \mathcal{N}$, $m\! \neq \! n$ and the noise $n[t]$. The simulation results are obtained after averaging over $100$ realizations.

We consider two metrics for performance evaluation: consensus error (CE) $e_{\mathrm{ce}}[t]$ and root mean square error (RMSE) $e_{\mathrm{rmse}}[t]$, which are defined as
\begin{align}
		\label{Eq.25}	
		& e_{\mathrm{ce}}[t] = \sqrt{\sum\nolimits_{n=1}^{N} (x_{n}[t] - x^{*})^{2}/N};  \\
		\label{Eq.26}	
		& e_{\mathrm{rmse}}[t] =  \sqrt{\sum\nolimits_{n=1}^{N} (x_{n}[t] - \sum\nolimits_{m=1}^{N} x_{m}[t] /N )^{2}/N}.
\end{align}

\begin{figure}[h!]
	\centering
	\subfloat[AC]{
		\includegraphics[width=0.22\textwidth]{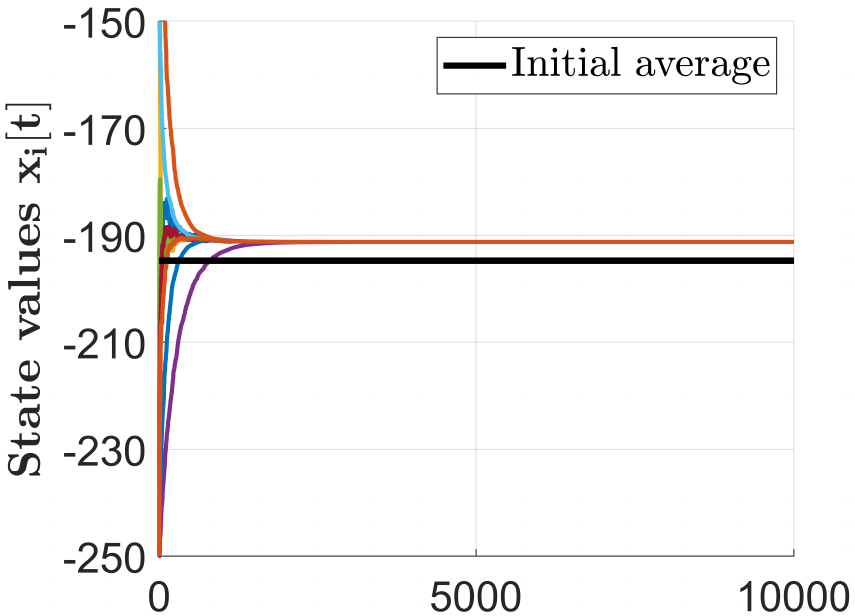}
		\label{Fig.2:sub1}
	}
	\hspace{0pt}
	\subfloat[AC-PCSS]{
		\includegraphics[width=0.22\textwidth]{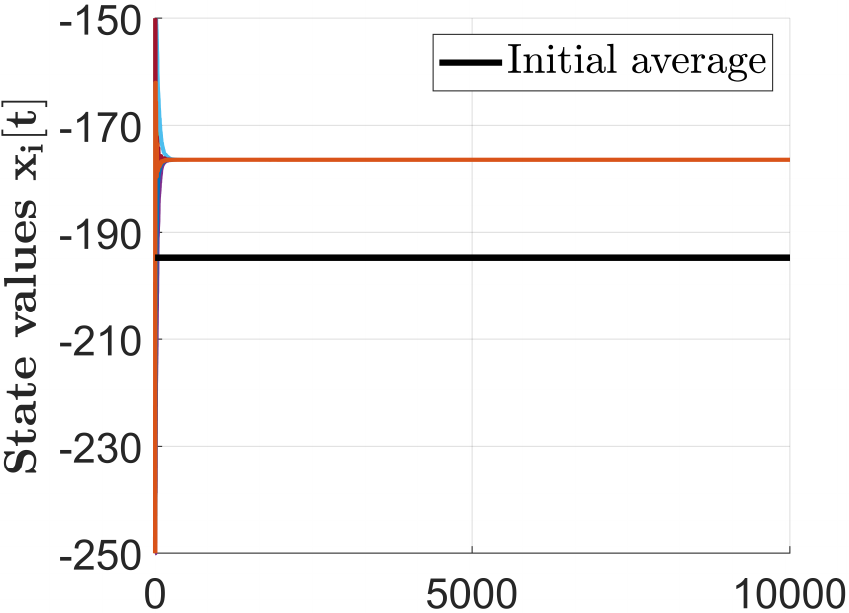}
		\label{Fig.2:sub2}
	} \\
	\subfloat[D-PGD-AC]{
		\includegraphics[width=0.22\textwidth]{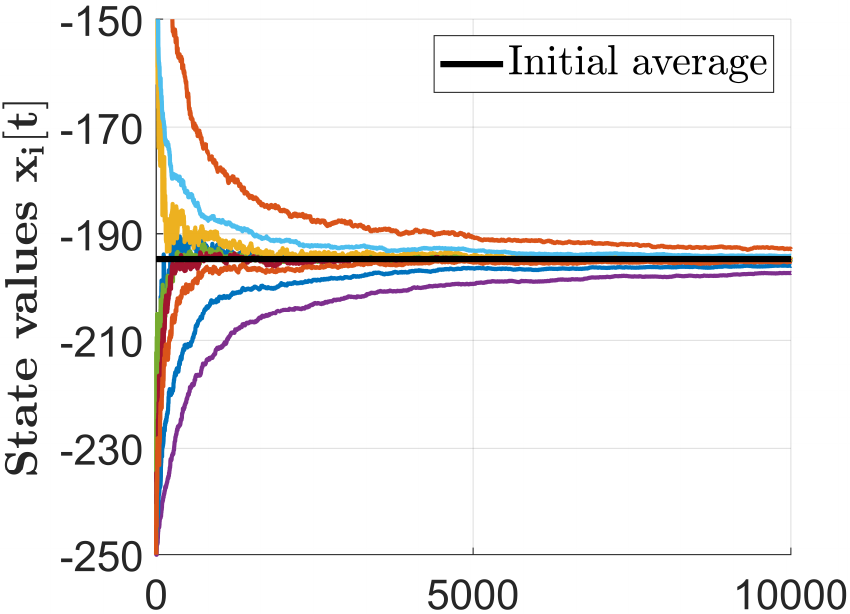}
		\label{Fig.2:sub3}
	}
	\hspace{0pt}
	\subfloat[D-PGD-AC-PCSS]{
		\includegraphics[width=0.22\textwidth]{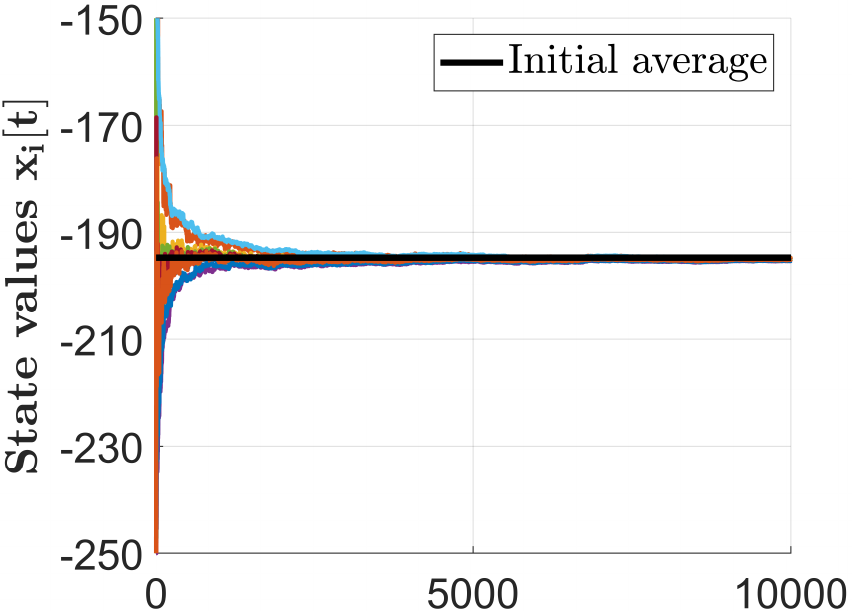}
		\label{Fig.2:sub4}
	}
	\caption{The evolution of $x_{n}[t]$ for 9 nodes in 10000 iterations.}
	\label{Fig.2}
\end{figure}

\subsection{Performance evaluation on consensus accuracy}

\begin{figure*}[h!]
	\centering
	\subfloat[]{
		\includegraphics[width=0.30\textwidth]{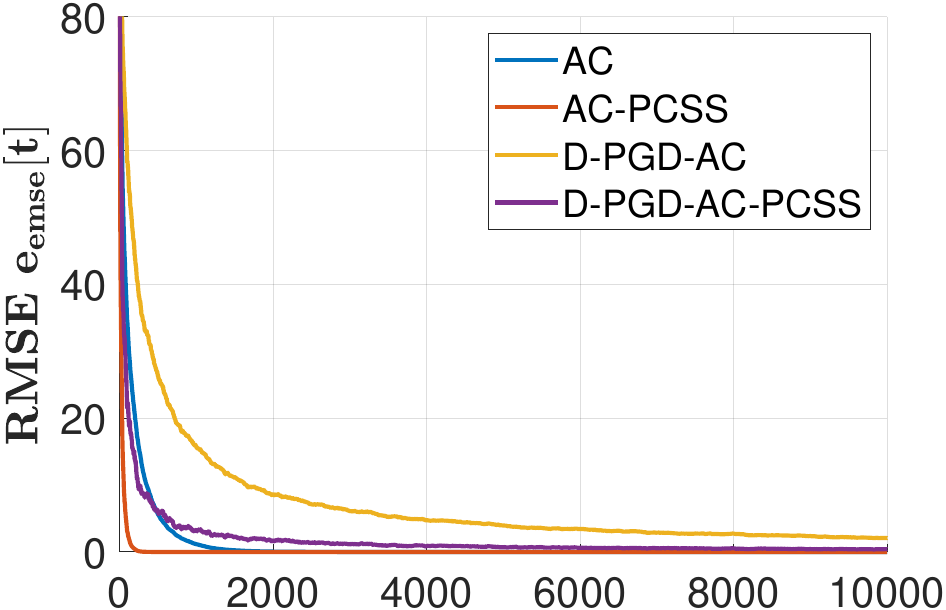}
		\label{Fig.3:sub1}
	}
	\hspace{0pt}
	\subfloat[]{
		\includegraphics[width=0.30\textwidth]{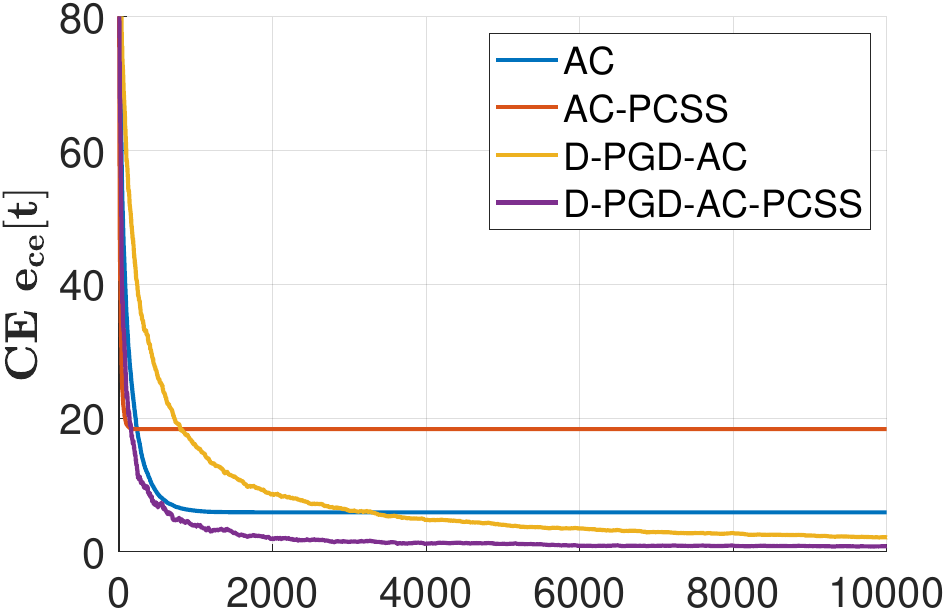}
		\label{Fig.3:sub2}
	}
	\caption{Convergence performance comparison. (a) Root mean square error (RMSE). (b) Consensus error (CE).}
	\label{Fig.3}
	\vspace{-3mm}
\end{figure*}

First, we evaluate the accuracy of the four algorithms by showing the evolution of $x_{n}[t]$ for all $n \in \mathcal{N}$. As illustrated in Fig. \ref{Fig.2}, both the AC and AC-PCSS algorithms converge to certain values that deviate from $x^*$. The existence of non-coherent interference ($c_{2,n}[t]$) and aggregated noise ($c_{3, n}[t]$ ) are the primary reasons causing this bias, because of which $\mathbf{W}[t]$ cannot fulfill the convergence conditions (C1) and (C2). Consequently, the average of initial state values is not preserved over the iterations.
Thanks to the decreasing step sizes $\zeta[t]$ and $\eta[t]$, $c_{2,n}[t]$ and $c_{3, n}[t]$ will vanish as $t \rightarrow \infty$ and agents will ultimately reach a consensus, although not on the exact average value $x^*$. Given that the expectation of $c_{2,n}[t]$ grows in polynomial order with $N$, the bias of AC and AC-PCSS algorithms will increase with $N$. In contrast, the D-PGD-AC and D-PGD-AC-PCSS algorithms can correct this deviation and achieve MSAC on $x^*$, showing their robustness against non-coherent interference and noise. Incorporating $\mathbf{d}[t]$ contributes to this robustness effect, since $\mathbf{d}[t]$ acts as a regularization term and directs the system towards $x^*$. 

Fig. \ref{Fig.3} compares the performance of the four algorithms in terms of $e_{\mathrm{ce}}[t]$ and $e_{\mathrm{rmse}}[t]$. Fig. \ref{Fig.3:sub1} demonstrates that all four algorithms can converge to a consensus value. However, Fig. \ref{Fig.3:sub2} reveals that the consensus values of the AC and AC-PCSS algorithms exhibit a bias relative to $x^*$. Specifically, these two algorithms converge after $t=2000$ and maintain a consistent deviation from $x^*$, whereas the consensus error in the D-PGD-AC and D-PGD-AC-PCSS algorithms continues to decrease. 
As $t \rightarrow \infty$, the latter two algorithms can converge to $x^*$.

\subsection{Performance evaluation on convergence rate}
Next, we compare the convergence rate of the four algorithms. From the results in Fig.~\ref{Fig.2}, we see that employing the PCSS method can significantly increase the convergence rate. However, as shown in Fig. \ref{Fig.3:sub2}, while the PCSS method increases the convergence rate of the AC-PCSS algorithm, it also enlarges its consensus error (derivation between converged value and true average of initial values $x^*$). 
This implies a trade-off between accuracy and convergence rate. When agents converge faster, they fail to obtain sufficient information from their neighbors.
Consequently, agents are more significantly influenced by interference and noise. In contrast, the PCSS method does not compromise convergence accuracy when applied to our proposed algorithm (D-PGD-AC-PCSS). This is because the D-PGD-AC-PCSS algorithm is guaranteed to converge to an unbiased estimation of $x^*$. Therefore, our D-PGD-AC-PCSS algorithm achieves both accurate and fast convergence to the true average of the initial state values.

\section{Conclusions}
\label{Conclusions}
In this work, we explore the use of non-coherent OTA aggregation in distributed average consensus systems. We propose two enhancement schemes to improve the robustness and convergence rate of existing non-coherent OTA average consensus algorithms. Specifically, we adopt a D-PGD-based average consensus algorithm with power control, which ensures mean square convergence to the exact consensus value (the average of initial state values) while accelerating the convergence rate. The benefits of our proposed design are validated through simulations, demonstrating that power control effectively modifies the topology of wireless multi-agent systems employing non-coherent OTA aggregation for distributed consensus, thereby enhancing convergence speed.

\bibliographystyle{IEEEtran}
\bibliography{ref}
	
\end{document}